\documentclass[letter]{aa} 
\usepackage{graphicx}
\usepackage{txfonts}
\usepackage{epsfig,color}
\def\4U{4U\thinspace1705-44 }
\begin{document}
%
\title{BeppoSAX observation of 4U\thinspace1705-44: detection of hard X-ray emission in the soft state}
  \subtitle{}
   \author{S.~Piraino \inst{1,2} \and A.~Santangelo\inst{2}
     \and T.~Di Salvo\inst{3} \and P.~Kaaret\inst{4}
     \and D.~Horns\inst{2} \and R.~Iaria\inst{3}\and L.~Burderi\inst{5}}


\institute{INAF-IASF di Palermo, Via Ugo La Malfa 153, 90146 Palermo, Italy\\
\email{Santina.Piraino@ifc.inaf.it} \and IAAT, University of T\"{u}bingen, Sand
1, 72076 T\"{u}bingen, Germany \and Dipartimento di Scienze Fisiche ed
Astronomiche, Universit\`a di Palermo, via Archirafi 36 - 90123 Palermo, Italy
\and Department of Physics and Astronomy, University of Iowa, Iowa City, IA
52246, USA \and Universit\'a degli Studi di Cagliari, Dipartimento di Fisica,
SP Monserrato-Sestu, KM 0.7, 09042 Monserrato, Italy}

   \date{Received ; accepted }

\abstract
{\4U is one of the best-studied type I X-ray burster and atoll sources.}
{Since it covers a wide range in luminosity (from a few to 50 $\times$
10$^{36}$ erg s$^{-1}$) and shows clear spectral state transitions, it
represents a good laboratory for testing the accretion models proposed for
atoll sources.}
{We analyzed the energy spectrum accumulated with  BeppoSAX observations (43.5
ksec) in August 2000 when the source was in
 a soft spectral state.}
{The continuum of the wide-band energy spectrum is well-described by the sum of
a blackbody (kT$_{bb}$$\sim$0.56~keV) and a Comptonized component (seed-photon
temperature kT$_W$$\sim$1~keV, electron temperature kT$_e$$\sim$2.7~keV, and
optical depth $\tau$$\sim$11). A hard tail was detected at energies above
$\sim$25~keV. The latter can be modeled by a power law having a photon index
$\sim$2.9, which contributes $\sim$11\% of the total flux in the range  0.1-200
keV. A broad emission line, possibly from a relativistic accretion disk, models
the feature in the Fe K line region of the spectrum.}
{This is the first time that a high-energy tail has been observed during a soft
state of the source.}

 \keywords{accretion, accretion disks -- stars: individual: 4U 1705-44--- stars:
neutron --- X-rays: stars --- X-rays: binaries--- X-rays: general}

   \authorrunning{Piraino et~al.\ 2007}
   \titlerunning{Hard X-ray emission from \4U}

   \maketitle

\section{Introduction}

Numerous similarities have been found between low-mass X-ray binaries
containing neutron stars and accreting stellar-mass black holes. The similarity
between the X-ray properties of these systems suggests that the fundamental
physical processes producing X-ray emission must be quite similar. Indeed,
neutron stars and black holes have very similar gravitational potentials, and
the neutron star radius is close to the size of the last stable orbit of
material around a black hole, so the properties of the accretion flow around
the compact object should be similar. On the other hand, since neutron stars
 have a solid surface, unlike black holes, they should show unique signatures.
X-ray bursts and coherent pulsations are unique signatures of neutron stars.
However, they are a sufficient, but not necessary condition, for the presence
of a solid surface, since not all neutron star systems show these features.

There have been attempt to look for effects of the stellar surface on the
energy spectrum by comparing black holes and neutron stars in similar
luminosity states.  The proposed black hole spectral features, such as an
ultra-soft spectral state at high luminosity, hard X-ray emission at low
luminosity, and steep X-ray tail seen at high luminosity, have been
contradicted by counter-examples found in neutron-star spectra (see e.g.\
\cite{Tanaka95}; \cite{Barret94}; \cite{DiSalvo01}). A unique black hole
spectral signature has remained so far elusive.

\cite{Hasinger89} divided the low-mass X-ray binaries containing neutron stars
into two groups called Z sources and atoll sources after the patterns these
sources trace out in the X--ray color-color diagram (CD). The CD of Z sources
displays a Z-like track, whereas the CD of atoll sources show a C-like track
where two branches can be identified, the island and the  banana states. The
island state is characterized by lower count rates, slower motion in CD and
stronger band-limited noise than the banana state. For both classes, the
position of the source in the CD is thought to be an indicator of the mass
accretion rate ($\dot{M}$) increasing from the upper left to the bottom right
of the Z track for Z sources and from the island state to the right of the
horizontally elongated banana state for atoll sources. The Z sources are
thought to have larger magnetic fields and/or higher $\dot{M}$ than atoll
sources.

Recently, it has been shown that after sampling the source intensity states of
atoll sources enough long, the shape of their CD tends to resemble those of Z
sources (\cite{Gierlinski02}; \cite{Muno02}). However, the similarities of the
atoll CD track with that of Z sources are mainly confined to the banana branch.
Moreover, the properties of atoll sources along the CD track are very different
from those shown by Z sources (e.g.\ \cite{Barret02}; \cite{Klis06}).

The bright low-mass X-ray binary \4U  was classified  by Hasinger \& van der
Klis (1989) as an atoll source. Long-term monitoring of this source succeeded
in sampling the complete track in the CD (\cite{Gierlinski02}; \cite{Muno02};
\cite{Barret02}).

\4U is an X-ray burster (\cite{Sztajno85}), whose bursting activity and
frequency depend on the persistent emission. During the low state, when type I
X--ray bursts are most frequent, the spectrum is unusually hard
(\cite{Langmeier87}).

In this Letter, we report on measurements with the  broad band coverage,
sensitivity, and good energy resolution of BeppoSAX (\cite{Boella97a})  of the
\4U X--ray spectrum obtained in the energy range 0.3--200~keV at the beginning
of a clear transition from a high  to a low flux level.

\section{Observations}
We performed a BeppoSAX observation of 4U\thinspace1705-44, on August 20--21,
2000 (MJDs=51766/7) for a total of 43.5~ksec of on--source observing time.
Here, we report on the results of the average energy spectrum. Details on the
timing and spectral evolution along the color-color diagram will be reported in
a subsequent paper (\cite{Piraino07}).

We obtained spectra from the four BeppoSAX Narrow Field Instruments (NFIs) in
energy bands selected to give a good signal-to-noise ratio for this source: the
Low Energy Concentrator Spectrometer (LECS, 0.3--4~keV; \cite{Parmar97}), the
Medium Energy Concentrator Spectrometer (MECS, 2--10~keV; \cite{Boella97b}),
the High Pressure Proportional Gas Scintillation Counter (HPGSPC, 8--50~keV;
\cite{Manzo97}), and the Phoswich Detection System (PDS, 15--200~keV;
\cite{Frontera97}). The LECS and MECS data were extracted in circular regions
centered on the source position using radii of 8' and 4', respectively,
containing 95\% of the source flux. Identical circular regions in blank field
data  were used for the extraction of background spectra and  background
subtraction. As the source lies in  the direction of the Galactic bulge, we
scaled the blank field background at the source position to the mean level of
the background around the source during the observation using a factor of 2.7
for LECS and 2.64 for MECS. Spectra accumulated from Dark Earth data and during
off--source intervals were used for the background subtraction for HPGSPC and
PDS spectra, respectively.

The LECS and MECS  spectra were rebinned to sample the instrument energy
resolution with the same number of channels at all energies, and the HPGSPC and
PDS spectra were grouped using a logarithmic grid. The spectral analysis was
performed with the data analysis package XSPEC v. 11.2 (\cite{Arnaud96}). A
normalization factor has been included to account for the mismatch in the
BeppoSAX instruments' absolute flux calibration. We found relative
normalizations in good agreement with typically observed values
(\cite{Fiore99}).
\section{Spectral analysis}
We fit the 0.3--200~keV  BeppoSAX spectrum of \4U with several single or
multicomponent continuum models. The combination of the {\tt comptt}
Comptonization model (\cite{Titarchuk94}) plus blackbody gave a fit having
$\chi^{2}/d.o.f$=964.3/530 ($\chi^2_\nu\simeq$1.82). Using a multicolor disk
blackbody, {\tt diskbb}, instead of a simple blackbody, we obtained a
$\chi^{2}/d.o.f=$1077/530  ($\chi^2_\nu\simeq$2.03).

The fit was improved by adding a broad-iron K-line feature. Using a simple
Gaussian line at 6.58$\pm$0.05 keV ($\sigma$=0.31$\pm$0.08 keV, EW=63 eV), the
fit gave a $\chi^{2}/d.o.f$=701.28/527 ($\chi^2_\nu\simeq$1.33). Line emission
from a relativistic accretion disk ({\tt diskline}, \cite{Fabian89}) at
$\simeq$6.7~keV works slightly better with $\chi^{2}/d.o.f$=678.75/526
($\chi^2_\nu\simeq$1.29). We report the best-fit parameters in Table~1.
\begin{table}[t]
\caption[]{Results of fitting the \4U spectra in the energy band 0.3--200~keV.}
\begin{center}
\label{tab1}
\begin{tabular}{l|c}
\hline \hline
  Parameter& Value       \\
 \hline
$N_{\rm H}$ $\rm (\times 10^{22}\;cm^{-2})$ &  $1.9 \pm 0.1$        \\
$k T_{\rm bb}$ (keV)                        & $0.56 \pm 0.01$        \\
$N_{\rm bb}$ (keV)                          & $(2.23 \pm 0.11)\times 10^{-2}$  \\
$k T_{\rm W}$ (keV)                         & $1.13 ^{+0.05}_{-0.02}$        \\
$k T_{\rm e}$ (keV)                         & $2.72 \pm 0.09$          \\
$\tau$                                      & $11.0 \pm 0.6$          \\
$N_c$                                       & $0.35 \pm 0.02$        \\
$E_{\rm Fe}$ (keV)                          & $6.7^{+0.2}_{-0.5}$          \\
$R_{\tt in}$ (R$_g$)                            & $8.1^{+4.2}_{-2.1}$          \\
$Incl$ (deg)                                & $28^{+20}_{-8}$       \\
$I_{\rm Fe}$ (ph cm$^{-2}$ s$^{-1}$)        & $(4.7^{+2.0}_{-0.6}) \times 10^{-3}$ \\
Fe $Eq. W.$ (eV)                            & 109                     \\
$\Gamma$                                    & $2.9^{+0.2}_{-0.3}$        \\
$N_p$                                       & $0.7 ^{+0.4}_{-0.2}$        \\
$F_{\rm tot ab}$                            & $5.81$                  \\
$F_{\rm tot}$                               & $16.7$  \\
$\chi^2_{\rm red}$/d.o.f.                 & 1.183 /524              \\
  \hline
\end{tabular}
\end{center}
{\small \sc Note} \footnotesize--- The model consists of {\tt blackbody},
Comptonization {\tt comptt}, and power-law components, and a {\tt diskline}
emission line, all subject to interstellar absorption. Uncertainties are at the
90\% confidence level for a single parameter. The total flux is in the
0.1--100~keV energy range and in units of $\times 10^{-9}$ ergs
cm$^{-2}$s$^{-1}$.
\end{table}
An excess of the counts above $\sim$25~keV was apparent in the spectrum. None
of the two-component continuum models we tried could model this excess. By
adding a power-law with a photon index 2.9 to the {\tt blackbody+comptt}
continuum model, the fit improved significantly ($\chi^{2}/d.o.f$=620/524,
$\chi^2_\nu\simeq$1.18; an F-test gave a probability of chance improvement of
the fit of $\simeq 10^{-10}$). This component contributed $\sim$11\% of the
0.1--200~keV source flux. The residuals after including a power-law component
are shown in the bottom panel of Fig.1. The average spectrum, together with the
total model and its components, is shown in Fig.2

Note that \cite{Fiocchi07} present a  spectral analysis of INTEGRAL and
BeppoSAX data of 4U 1705-44, which includes this BeppoSAX observation, claiming
the detection of a Compton reflection component. However they did not include
HPGSPC data in their analysis. Using their model in the whole BeppoSAX
(0.1--200 keV) range, we obtain a $\chi^2/d.o.f.$=715.9/526 and evident
residuals in the 10--50 keV range. We therefore conclude that the model
presented here describes the energy spectrum of the source better in the entire
BeppoSAX band.

\begin{figure}[t]
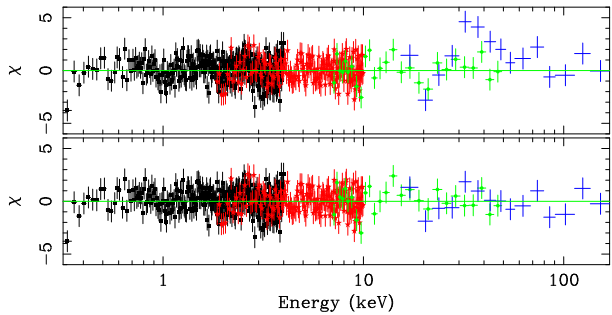

\hbox{\psfig{file=7841f1u.eps,height=8.0cm,angle=270.0}}
{\psfig{file=7841f1b.eps,height=8.0cm,angle=270.0}} {\caption[]{Residuals in
units of $\sigma$ with respect to the best-fit model without (upper panel) and
with (bottom panel) the power-law component. } \label{res}}
\end{figure}
\begin{figure}[h]
\hbox{\psfig{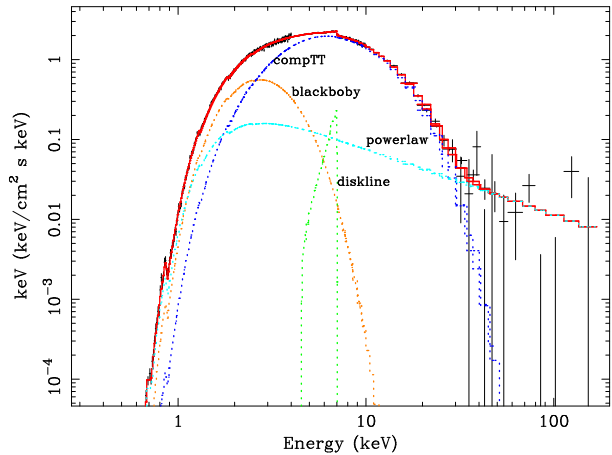}} {\caption[]{X-ray
(unfolded) spectrum together with the best-fit model (red line) and its
components: blackbody (orange), Comptonization (blue), power-law (blue-green),
diskline (green).} \label{spectrum}}
\end{figure}

\section{Discussion}
We have analyzed  a BeppoSAX observation of the atoll source and X-ray burster
4U\thinspace1705-44. The observation was performed in August 2000 for a total
exposure time of 43.5 ksec. A broad-band (0.3--200~keV) spectral analysis
reveals that the source continuum spectrum is well-fitted by a {\tt blackbody}
($kT_{BB}\simeq0.56$ keV), contributing $\sim$14\% of the observed 0.1--200~keV
flux, plus a Comptonized component modeled by {\tt comptt} (seed photon
temperature of $\sim$1.1~keV, electron temperature $\sim$2.7~keV, and optical
depth $\sim$11), plus a hard power-law component (photon index $\sim$2.9)
contributing about 11\% of the 0.1--200~keV source flux. An emission line at
6.4--6.7~keV is also significantly detected. This feature is quite broad and,
in line with previous works (\cite{DiSalvo05}), we fitted it with a {\tt
diskline} model, yielding an inner disk radius of $\sim$$8\;$ $R_g$ ($R_g = G M
/ c^2 = 1.5$~km for $M = 1 M_\odot$ is the gravitational radius) and a
relatively low inclination of about $30^\circ$. The 0.1--200 ~keV source
luminosity was 3.8 $\times$ 10$^{37}$ erg s$^{-1}$ assuming a distance to the
source of 7.4 kpc (Haberl \& Titarchuk 1995).

The persistent emission spectrum of \4U has been discussed by
Barret \& Olive (2002) and recently by Di Salvo et al. (2005).
Barret \& Olive (2002) observed \4U with  the Rossi X-Ray Timing
Explorer during a clear spectral state transition (soft-hard-soft)
and interpreted the observed spectral evolution of the source
within a scenario in which the accretion geometry is made of a
truncated accretion disk of varying radius and an inner flow
merging smoothly with the neutron star boundary layer
(\cite{Barret00}; \cite{Done02}). Barret \& Olive (2002) show that
the truncation radius is not set by the instantaneous accretion
rate, as observations with the same bolometric luminosity may have
very different spectral and timing properties. A possible
mechanism for the state transition is the evaporation of the disk
due to the conduction of heat between the hot inner flow and the
cold disk (\cite{Rozanska00}). The disk is a powerful source of
cooling for the Comptonization; the closer the disk gets to the
neutron star, the more effectively it cools the inner flow (Done
2002). In agreement with Barret \& Olive (2002), we interpret the
soft ({\tt blackbody}) component observed during the BeppoSAX
observation as dominated by the disk emission at the highest
luminosity (a contribution from the neutron star surface cannot be
excluded). The fact that a simple blackbody provides a better fit
to the soft component of the spectrum than the usual diskbb model
may suggest that the disk differs from a standard thin disk
because of the presence of a boundary layer. The Comptonized
component probably originates from the (hotter) inner flow, since
the seed photon and electron temperature of our best-fit model are
higher than the temperature of the blackbody component detected in
the spectrum.  The high optical depth of the Comptonization
component, $\tau$$\sim$11, indicates that the source of the seed
photons should lie inside the Comptonization region and not be
directly visible as a component in the X-ray spectrum.

Di Salvo et al.\ (2005) observed \4U  in the energy range 0.3--10 keV, with
Chandra during a soft state and modeled the continuum of the energy spectrum
with a soft Comptonization model, {\tt comptt} (electron temperature
kT$_e$$\sim$2.3 keV and optical depth $\tau$$\sim$18 for a spherical geometry).
The (0.1--10 keV) luminosity was 3.3 $\times$ 10$^{37}$ erg s$^{-1}$. Both the
spectral parameters and the X-ray luminosity of the source during the Chandra
observation are similar to the spectral parameters and luminosity during our
BeppoSAX observation, suggesting that the source in both cases was in a soft
spectral state. In particular, we probably observed  the source in a soft/high
spectral state in the lower banana state of the atoll-track. In the RXTE-ASM
energy range, this observation was at the beginning of a transition between
high rate and low rates and at a flux level intermediate between the ones
reported by Di Salvo et al. (2005) and Barret \& Olive (2002) (Fig.3).

\begin{figure}[h]
\hbox{\psfig{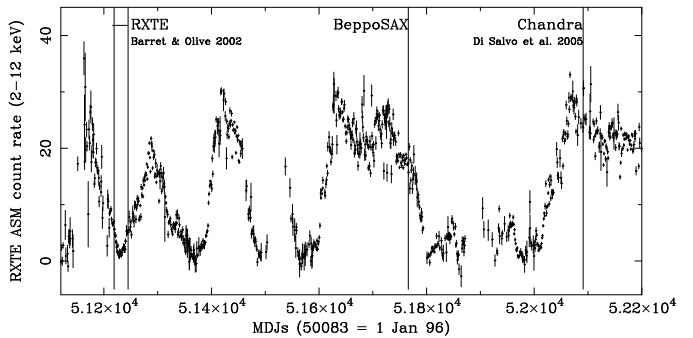}} {\caption[]{Long-term
RXTE/ASM 2-12 keV light curve (HEASERC public data base).  The time of the
BeppoSAX, RXTE, and Chandra observations are indicated with vertical lines.}
\label{lc}}
\end{figure}

Three discrete features (residuals from the continuum model fits) at energies
below 3~keV were fitted and identified in the Chandra X-ray spectrum of
4U\thinspace1705-44. The iron $K\alpha$ line at 6.5~keV was found to be
intrinsically broad (FWHM $\sim$1.2~keV) and compatible with reflection from a
cold accretion disk with $R_{in}$ $\sim$15~km or with a Compton broadening in
the external parts of a $\sim$2 keV corona. A broad and strong emission line is
also necessary to fit the BeppoSAX spectrum. In agreement with Di Salvo et al.\
(2005), we fit this feature with a {\tt diskline} model, finding best-fit
values of the spectral parameters that agree well with the Chandra results. In
particular we find the line centroid energy at $6.7^{+0.2}_{-0.5}$~keV,
compatible, within the uncertainty of the BeppoSAX spectral fitting, with the
line energy of 6.4~keV found from the Chandra observation. The main difference
between our results and the Chandra results is in the best-fit value of the
disk inclination, which is higher for Chandra (59$^{+25}_{-4}$ deg) than for
BeppoSAX. Indeed, if we fix the line centroid energy to $6.4$ keV, the best-fit
value from the Chandra observation, we find an inclination angle of the disk
between $20^\circ$ and $50^\circ$, barely compatible with the Chandra result.

In this Letter, we report, for the first time, the presence of a hard X-ray
(power-law shaped) component in the soft state of 4U\thinspace1705-44, with a
photon index of $\sim$ 2.9 contributing $\sim$11\% of the observed source flux.
Hard-energy tails, steep power-laws with photon indexes $\sim$2-3 and without
evidence of a high energy cutoff are observed in the spectra of accreting black
hole candidates (BHCs) during the intermediate state (IS) or very high state
(VHS).
 \footnote[1]{Note that \cite{Remillard&McClintock06} call the VHS state
the steep power law (SPL) state}.
 This component is variable and usually contributes a few per cent of the total flux in
the IS state, while in the VHS (or SPL) state it contributes up to 40--90$\%$
of the total flux (\cite{Remillard&McClintock06}). On the other hand, recent
broad band studies have shown that the spectra of most Z-sources display a
variable hard X-ray component (e.g.\ Di Salvo et~al.\ 2000, 2001, 2002;
\cite{Paizis06}). In some cases, it has been shown that the presence of this
hard component is correlated with the source spectral state. Specifically,  the
power-law component is the strongest at the lowest inferred mass accretion
rates. For the highest mass accretion rates of Z-sources (for instance in the
flaring branch of the CD) and in the high state, HS, of BHCs, the intensity of
the hard X-ray tail is often below detectability.

This suggests a similarity between the spectrum of the atoll source \4U  during
this soft banana state and the spectra of BHCs and Z-sources during some
spectral states, such as IS/VHS and horizontal and perhaps normal branch,
respectively. Indeed, other LMXBs of the atoll class have shown the presence of
a similar hard X-ray component (see e.g.\ \cite{Paizis06} for the bright atoll
source GX~13+1; \cite{Fiocchi06} for 4U~1636--53; \cite{Tarana07} for
4U~1820-30). We conclude that these hard X-ray components are becoming more and
more ubiquitous among X-ray binaries, indicating that similar emission
mechanisms and geometry characterizes all these systems. As proposed for
Z-sources, the hard tails in the soft state of \4U can be produced either in a
hybrid thermal/non thermal model (\cite{Poutanen98}) or in a bulk motion of
matter close to the neutron star (e.g. \cite{Titarchuk98}). Alternative
mechanisms that have been proposed is Comptonization of seed photons by
relativistic electrons in a jet (e.g \cite{DiSalvo00}) and synchrotron emission
from a relativistic jet escaping the system (\cite{Markoff01}).

\begin{acknowledgements}
This work was partially supported by the German Space Agency (DLR) under
contracts nos. 50 OG 9601 and 50 OG 0501, by the Consiglio Nazionale delle
Ricerche and by the Ministero della Istruzione, della Universit\'a e della
Ricerca (MIUR).
\end{acknowledgements}

\end{document}